\documentclass[aps,prl,twocolumn,showpacs,superscriptaddress]{revtex4}

\usepackage{graphicx}

\newcommand{\rb}{{\bf r}}

\begin{document}

\title{Resonant Correlation-Induced Optical Bistability in an Electron System on Liquid Helium}

\author{Denis Konstantinov}
\email[E-mail: ]{konstantinov@riken.jp}
\affiliation{Low Temperature Physics Laboratory, RIKEN, Hirosawa 2-1, Wako 351-0198, Japan}
\author{M. I. Dykman}
\affiliation{Department of Physics and Astronomy, Michigan State University, MI 48824, USA}
\author{M. J. Lea}
\affiliation{Department of Physics, Royal Holloway, University of London, TW20 0EX, UK}
\author{Yuriy Monarkha}
\affiliation{Institute for Low Temperature Physics and Engineering, 47 Lenin Avenue, Kharkov 61103, Ukraine}
\affiliation{Low Temperature Physics Laboratory, RIKEN, Hirosawa 2-1, Wako 351-0198, Japan}
\author{Kimitoshi Kono}
\affiliation{Low Temperature Physics Laboratory, RIKEN, Hirosawa 2-1, Wako 351-0198, Japan}

\date{\today}

\begin{abstract}
We show that electrons on liquid helium display intrinsic bistability of resonant inter-subband absorption. The bistability occurs for comparatively weak microwave power. The underlying giant nonlinearity of the many-electron response results from the interplay of the strong short-range electron correlations, the long relaxation time, and the multi-subband character of the electron energy spectrum.
\end{abstract}

\pacs{73.21.-b, 73.63.Hs, 03.67.Lx, 42.65.Pc }

\maketitle

Electrons on liquid helium provide a unique tool for studying correlation effects in two-dimensions (2D). The ratio of the characteristic Coulomb energy to the in-plane kinetic energy, the plasma parameter $\Gamma=e^2(\pi n_s)^{1/2}/k_B T$ ($n_s$ is the electron surface density), can vary by orders of magnitude, from $\Gamma\ll 1 $ where the electron system is a weakly-interacting gas to $\Gamma > 130$ where it is a Wigner crystal \cite{Grimes1979, Andrei}. In the broad range $1 < \Gamma <130$ the electron system is a correlated 2D electron liquid, with unusual and sometimes counter-intuitive properties of classical and quantum 2D magneto-transport \cite{Dykman1979} and activated and tunneling escape from the surface \cite{Iye1980}.

The effect of correlations in an electron liquid was also seen in weak-field spectroscopy of transitions to excited subbands of motion along the helium surface \cite{Lambert1980}. In a sense, it is a counter-part, for a strongly correlated system, of the depolarization effect in inter-subband absorption in semiconductor heterostructures with high electron densities \cite{Allen1976}. For semiconductors, of much interest are both correlation transport effects \cite{Spivak2009} and nonlinear optical effects related to radiation-induced population of excited subbands \cite{Zaluzny1991}. Long sought has been optical bistability in intersubband absorption \cite{Stockman1993}.

For electrons on helium, optical nonlinearity should be strong, since the electron relaxation time is unusually long, reaching $\sim 10^{-7}$~s for $T\gtrsim 0.1$~K, and saturation of resonant inter-subband absorption has been indeed seen \cite{Collin2002}. The interplay of the long relaxation time and strong spatial correlations provides a qualitatively new nonlinearity mechanism and should lead to new resonant effects. Such effects are indeed found in this paper.

Our central result is the first, to the best of our knowledge, direct experimental observation and a theory of the bistability of resonant inter-subband absorption in a correlated electron system. The bistability is due to the correlation-induced strong dependence of the inter-subband transition frequencies of an electron on the state of other electrons. An additional interest in this dependence comes from the proposals of quantum computing with electrons on helium \cite{Platzman1999}, as it provides a mechanism of two-qubit gate operations.

We study electrons on liquid $^3$He. The experimental setup is similar to that previously described~\cite{Konst2007}. Electrons are confined on a helium surface between two circular parallel electrodes in an asymmetric potential formed by the barrier on the surface, the image force, and an electric field $E_{\perp}$ normal to the surface from the voltage on the electrodes \cite{Crandall1972}. The quantized energies of 1D motion in this potential $\epsilon_l$ ($l=1,2\ldots$) give the inter-subband transition frequencies $\omega_{l'l}=(\epsilon_{l'}-\epsilon_l)/\hbar$ in the single-electron approximation. We resonantly excite $1\to 2$ inter-subband transitions with microwave (MW) radiation ($\omega/2\pi=104.5$~GHz) by tuning the frequency $\omega_{21}$ with the field $E_{\perp}$ through the linear Stark shift.

The resonant MW response is observed by measuring the low-frequency (100~kHz) in-plane magnetoconductivity $\sigma_{xx}$, which depends on the electron temperature $T_e$. We apply a classically strong magnetic field $B\sim 350$~G normal to the surface and use a Corbino disk that constitutes the top electrode. In these conditions, because of the strong electron correlations, $\sigma_{xx}\approx e^2n_s\nu/(m\omega_c^2)$, where $\omega_c=eB/mc$ and $\nu$ is the single-electron scattering rate with no magnetic field \cite{Dykman1979}. This rate, in turn, is determined by $T_e$ \cite{Saitoh1978}. Thus the variation of $\sigma_{xx}$ reflects the degree of electron heating caused by resonant MW absorption and allows us to determine $T_e$ from the relative change of the electron scattering rate $\nu$~\cite{Konst2007}.
\begin{figure}[b]
\centering
\includegraphics[width=7.5cm]{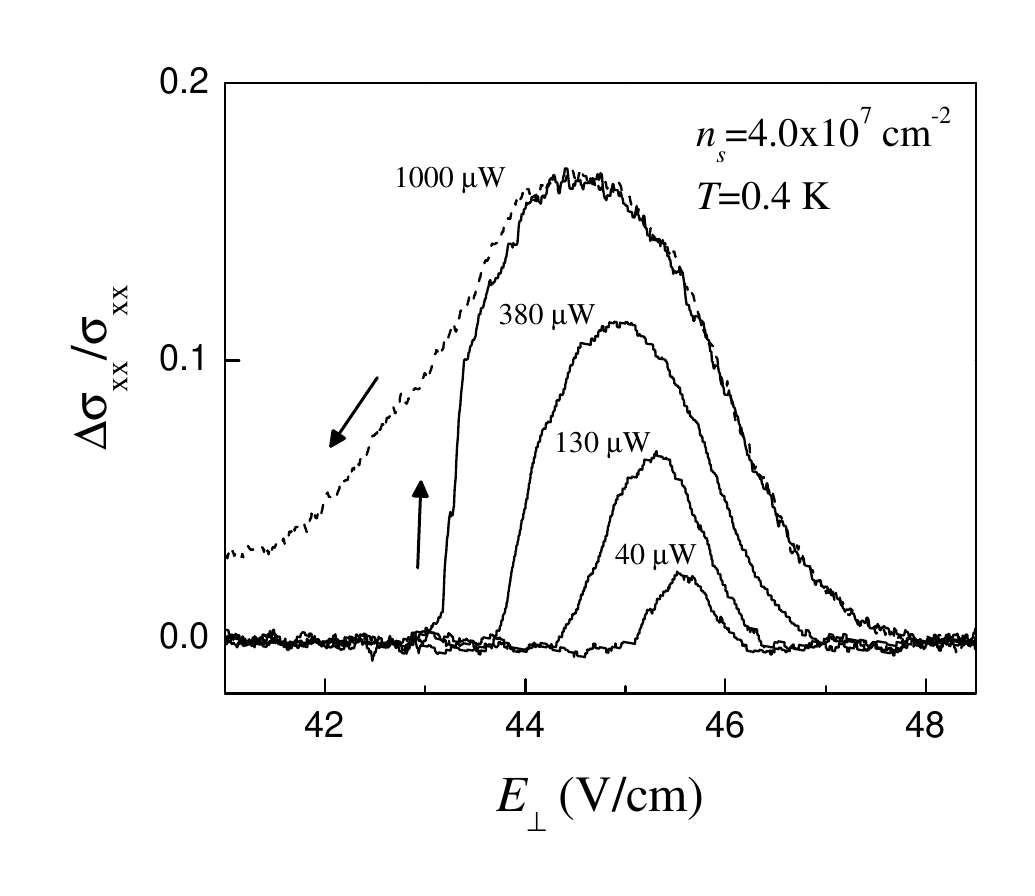}
\caption{\label{fig:1} Conductivity change $\Delta\sigma_{xx}/\sigma_{xx}$ vs. the tuning field $E_{\perp}$ for the surface density $n_s=4.0\times 10^7$~cm$^{-2}$ at $T=0.4$~K. Solid lines are traces obtained in an upward sweep. Corresponding to their increasing amplitude, the traces were taken at MW power $P=$~40, 130, 380 and $1000$~$\mu$W. The dashed line is the trace taken in a downward sweep at $P=1000$~$\mu$W (the sweeping directions are indicated by arrows).}
\end{figure}

The underlying physical picture is based on the hierarchy of relaxation processes in the resonantly modulated correlated electron system \cite{Ryvkine}. The fastest process is the exchange of in-plane momentum between the electrons. It occurs over the typical time $\omega_p^{-1}$, where $\omega_p=(2\pi e^2n_s^{3/2}/m)^{1/2}$ is the electron plasma frequency. The associated exchange of in-plane energy leads to establishing an in-plane electron temperature $T_e$, which is the same for all occupied subbands. Next comes the rate $\nu$ of quasi-elastic momentum scattering by helium vapor atoms or by surface waves on helium, ripplons. This scattering is short-range, and therefore $\nu$ characterizes not only the in-plane momentum relaxation rate, but also the rate of elastic inter-subband transitions. Such transitions lead to thermal distribution over subbands characterized by the same temperature $T_e$. The slowest process is the electron energy relaxation, which is due to inelastic processes, including one- and two-ripplon scattering or scattering by bulk excitations in helium \cite{Platzman1999}.

According to the above picture, when a MW field resonantly excites $1\to 2$ transitions, the energy goes first to heating up the electron system. As a result electrons populate higher subbands and the lowest subband is depleted, which leads to the decrease of resonant absorption. Such absorption bleaching occurs for much smaller radiation intensity than necessary for absorption saturation in a two-subband model \cite{Ryvkine,Konst2007}.

Figure~\ref{fig:1} shows the examples of the measured resonant response for $n_s=4.0\times 10^7$~cm$^{-2}$ at $T=0.4$~K, plotted as $\Delta\sigma_{xx}/\sigma_{xx}$, where $\Delta\sigma_{xx}$ is the change of the electron conductivity when MW radiation is applied. Each trace was obtained by slowly increasing $E_{\perp}$ to drive electrons through resonance, while maintaining the input radiation power $P$ at a fixed level. At $T=0.4$~K the scattering rate $\nu$ is dominated by collisions with He vapor atoms and monotonically increases with $T_e$~\cite{Saitoh1978,	Konst2007}. Correspondingly, upon sweeping $E_{\perp}$, $\sigma_{xx}$ increases and attains maximum at the resonance.

\begin{figure}[h]
\centering
\includegraphics[width=7.5cm]{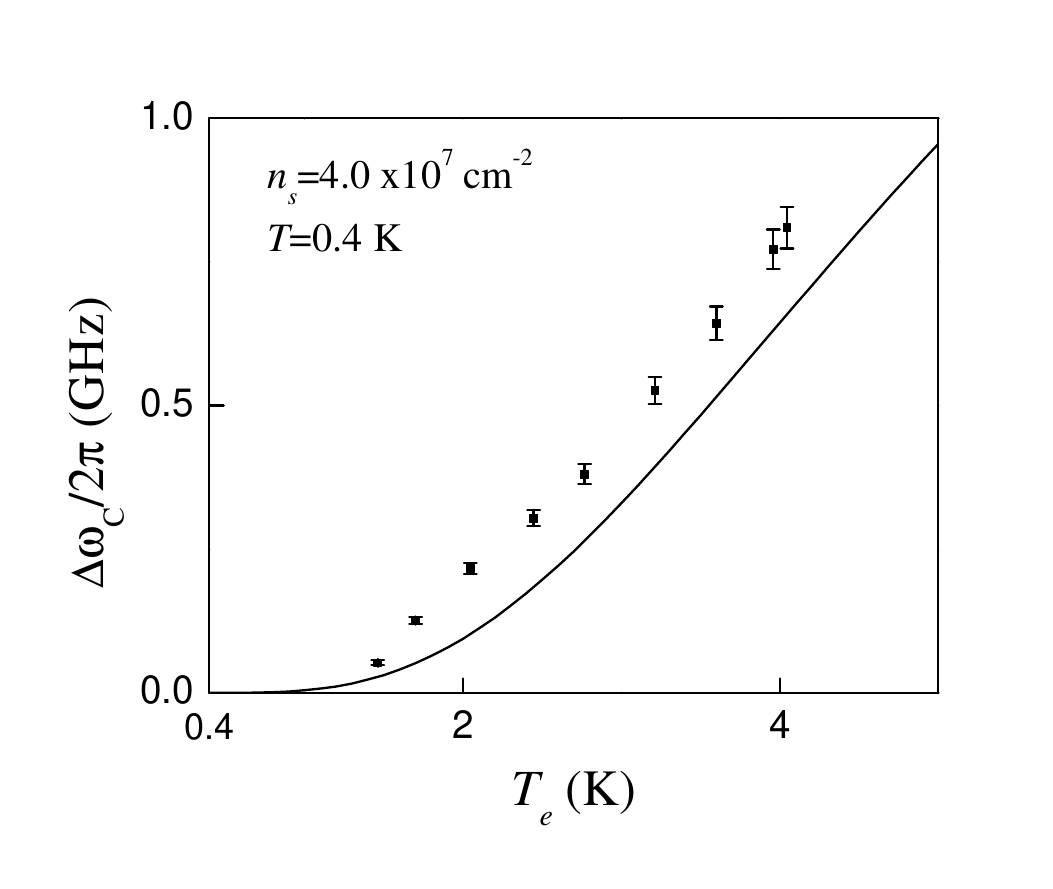}
\caption{\label{fig:2} Frequency shift $\Delta\omega_{\rm C}/2\pi$ vs. electron temperature $T_e$ obtained from the measured resonant response (squares). The solid line is given by Eq.~(\ref{eq:shift}) with $F$=8.91.}
\end{figure}

It is seen from Fig.~\ref{fig:1} that as the power increases, the resonance shifts towards lower field values. This means that the transition frequency $\omega_{21}$ increases relative to its value for electrons at $T_e\approx T$. The frequency shift $\Delta\omega_C$ obtained from the experiment is plotted in Fig.~\ref{fig:2} as a function of $T_e$. The conversion factor to frequency (the slope of $\omega_{21}/2\pi$ versus $E_{\perp}$) was determined experimentally by varying the frequency of the MW source and recording the corresponding shift of the resonance along the $E_{\perp}$ axis. This factor was found to be $0.64\pm 0.03$~GHz$\cdot $cm/V for the resonance in Fig.~\ref{fig:1}.

The major cause of the experimental frequency shift is the electron-electron interaction. The corresponding part of the Hamiltonian in a correlated system is
\begin{equation}
H_{ee}^{(z)}=-\frac{1}{4}e^2\sum\nolimits_{n\neq m} \left( z_n-z_m \right)^2/|{\textbf r}_n-{\textbf r}_m|^3,
\label{eq:ee}
\end{equation}
where ${\textbf r}_n $ and $z_n$ are the lateral and normal to the helium surface electron coordinates, respectively. We have taken into account that $|z_n-z_m|\ll |{\textbf r}_n -{\textbf r}_m|$. The distance $|z_n-z_m|$ is given by the localization length of the quantized states $|l\rangle$ of motion normal to the surface, which is $\sim 10$~nm \cite{Crandall1972}, whereas the inter-electron distance is $\gtrsim 1\,\mu$m for typical $n_s$ and $T_e$; note that $\langle H_{ee}^{(z)}\rangle$ would diverge if there were no spatial electron correlations.

Assuming that the lateral positions of the electrons and their distribution over the subbands are uncorrelated, to first order in $H_{ee}^{(z)}$ one obtains for the frequency shift \cite{Ryvkine}
\begin{eqnarray}
\label{eq:shift}
&&\Delta\omega_{21}=\frac{Fe^2n_s^{3/2}}{2\hbar} \Big[ (z^2)_{11}-(z^2)_{22} \nonumber\\
&&- 2\big( z_{11}- z_{22}\big)\sum\nolimits_l z_{ll}\rho_{ll} + 2|z_{12}|^2 \big( \rho_{11}-\rho_{22}\big)\Big],
\end{eqnarray}
where $A_{ll'}=\langle l|A|l'\rangle$; the matrix element $\rho_{ll}$ is the fractional occupation of subband $l$, and $F$ is related to the electron pair-correlation function $g(r)$ as $F=n_s^{-1/2}\int_{r=0}^{\infty} r^{-3}g(r)d\rb$. Equation (\ref{eq:shift}) differs from the corresponding expression in Ref.~\cite{Lambert1980} even for $\rho_{ll}=\delta_{l,1}$, as assumed there. Evidence of a transition frequency shift that depends on subband occupations was found by Glasson {\it et al.}~\cite{Glasson2004}.

If the many-electron system is described by the Boltzmann distribution with temperature $T_e$, the transition frequency shift, Eq.~(\ref{eq:shift}), is a function of $T_e$. The dependence of $\Delta\omega_{21}$ on $T_e$ comes from the occupations $\rho_{ll}$ and the factor $F$ that is determined by the plasma parameter $\Gamma$ calculated for $T=T_e$. For an electron crystal with triangular lattice $F\approx 8.91$. The experiment~\cite{Lambert1980} showed that $F$ does not increase significantly for $10\lesssim\Gamma\lesssim 100$. This agrees with the Monte-Carlo simulations \cite{Fang-Yen1997}. For experimental parameters $n_s$ and $T$ of Fig.~\ref{fig:1}, we have $\Gamma\sim 50$ for $T_e\approx T$. Therefore, to estimate $\Delta\omega_{21}$ we used $F=8.91$. The result for the electron-heating induced frequency shift $\Delta\omega_C = \Delta\omega_{21}(T_e) - \Delta\omega_{21}(T)$ is shown by the solid line in Fig.~\ref{fig:2}. The observed shift is larger by about 30$\%$ than the theoretical estimate. 
\begin{figure}[b]
\centering
\includegraphics[width=7.5cm]{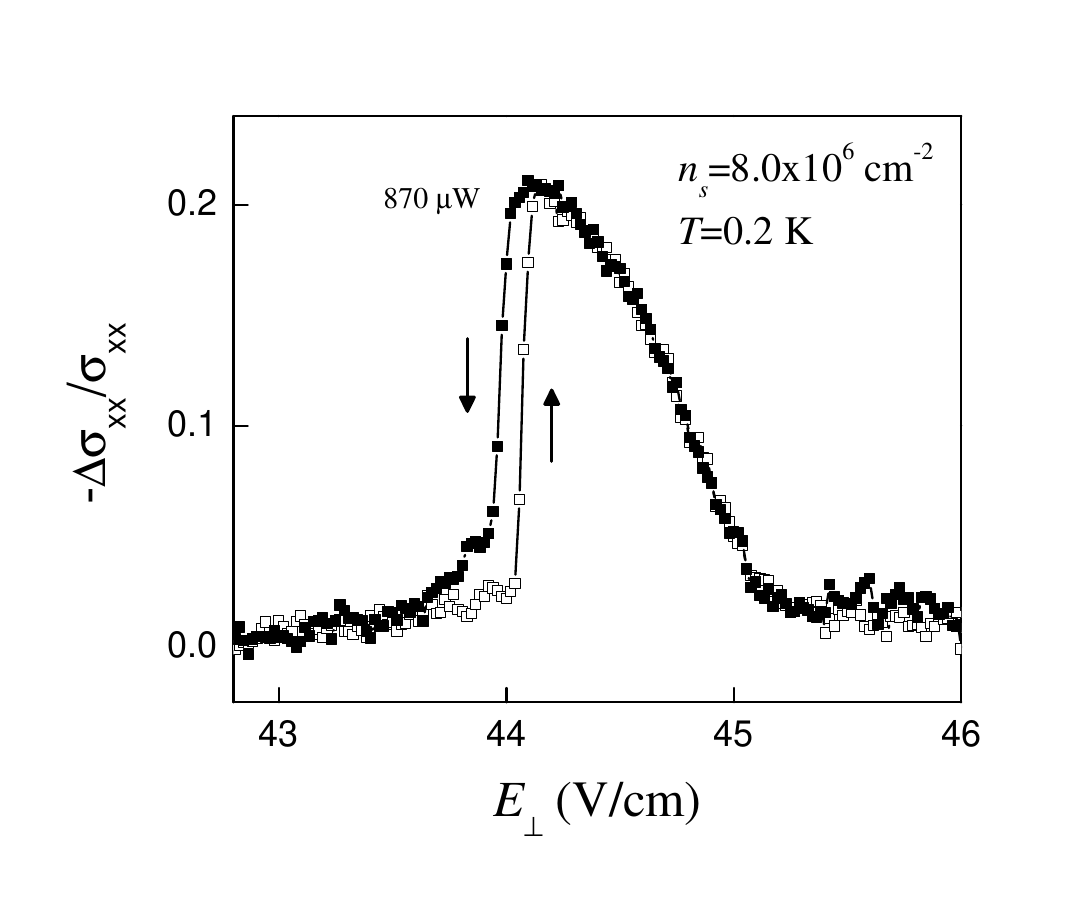}
\caption{\label{fig:3} Resonant response for $n_s=8.0\times 10^6$~cm$^{-2}$ at $T=0.2$~K obtained in an upward (open squares) and downward (solid squares) sweeps with $P=870$~$\mu$W. The sweeping directions are indicated by arrows.}
\end{figure}

The electron-electron interaction also affects the line shape of the resonance as seen in Fig.~\ref{fig:1}. As the MW power $P$ increases, the resonance line becomes asymmetric with a steeper low-field side. At high powers the line shape changes dramatically, as illustrated in Fig.~\ref{fig:1} by experimental traces taken at $P=1000$~$\mu$W. In an upward sweep, the low-field side becomes an abrupt jump that corresponds to a sudden increase of $T_e$. In a downward sweep, the response coincides with the upward-sweep curve on the high-field side of the resonance. However, it takes on substantially higher values on the low-field side. This unusual behavior becomes even more pronounced at lower temperatures. Figure~\ref{fig:3} shows a high power response for $n_s=8.0\times 10^6$~cm$^{-2}$ and $T=0.2$~K. At this temperature, the scattering rate $\nu$ (therefore $\sigma_{xx}$) is determined by the interaction of electrons with ripplons and decreases with $T_e$~\cite{Konst2007}. Correspondingly, the response is plotted as $-\Delta\sigma_{xx}/\sigma_{xx}$.

The hysteresis of resonant response indicate the onset of bistability, i.e. co-existence of two stable regimes characterized by different values of $T_e$. This effect can be explained by considering the balance between the heating of electrons by the MW radiation and cooling by the thermal bath (liquid helium). The energy balance equation can be written as~\cite{Ryvkine,Konst2007}
\begin{equation}
\hbar\omega r \left( \rho_{11} - \rho_{22} \right) = \nu_Ek_B\left( T_e -T \right).
\label{eq:bal}
\end{equation}
Here, $\nu_E$ is the electron energy relaxation rate  and $r$ is the rate of MW-induced transitions, $r=0.5\Omega_R^2\gamma/\left[(\delta+\Delta\omega_{21})^2+\gamma^2\right]$, where $\Omega_R=eE_0z_{12}/\hbar$ is the Rabi frequency ($E_0$ is the MW electric field), $\gamma$ is the half-width of the absorption line, and $\delta = \omega_{21}-\omega$ is (minus) the MW frequency detuning from the single-electron resonance.

\begin{figure}[h]
\centering
\includegraphics[width=7.5cm]{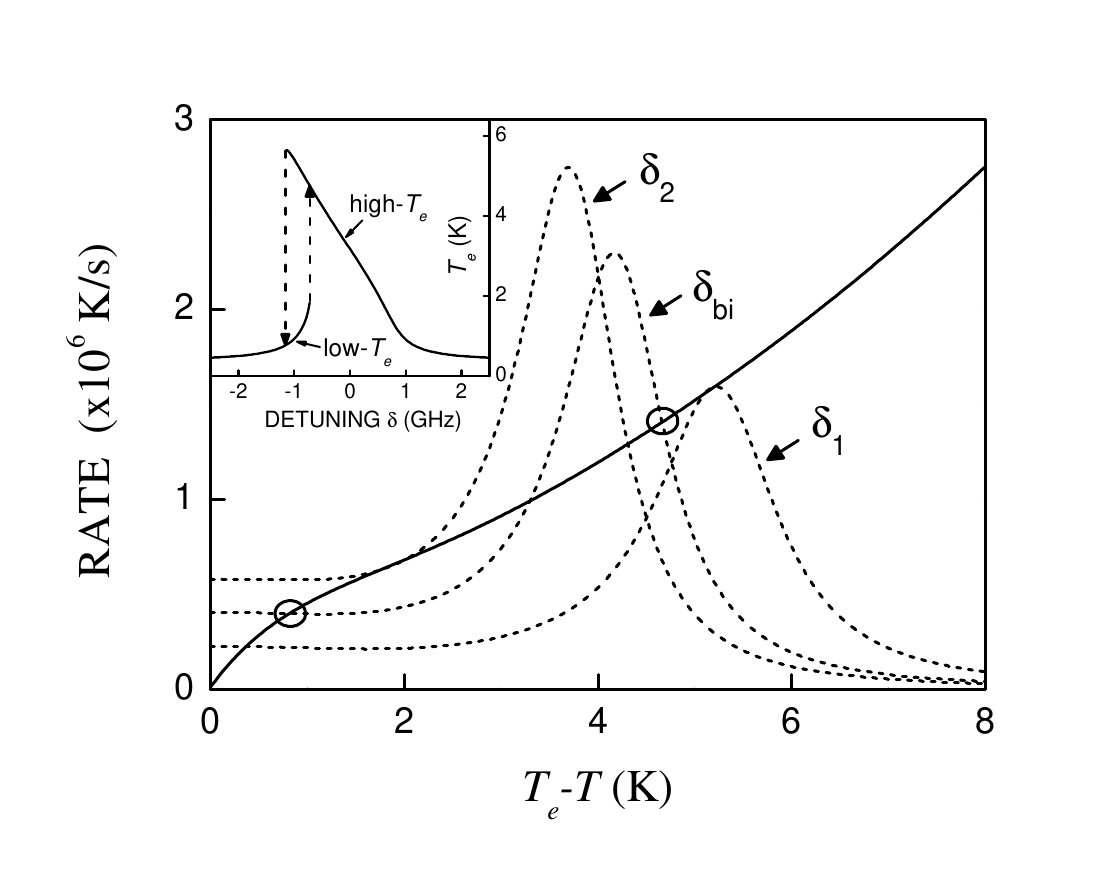}
\caption{\label{fig:4} Graphical solutions of the energy balance equation (\ref{eq:bal}) for $n_s=4.0\times 10^7$~cm$^{-2}$ and $T=0.4$~K. The solid line shows the cooling rate $\nu_E k_B (T_e-T)$. The dashed lines show the heating rate $\hbar\omega r(\rho_{11}-\rho_{22})$  calculated for three frequency detunings $\omega_{21}-\omega =\delta_1,\delta_{\rm bi}$ and $\delta_2$ (-1.15, -0.85 and -0.7~GHz, respectively). The other parameters are $\Omega_R/2\pi=0.01$~GHz, $\gamma/2\pi=0.2$~GHz and $\omega/2\pi=104.5$~GHz. The circles mark two stable solutions of Eq.~(\ref{eq:bal}) for detuning $\delta_{\rm bi}$. Inset: $T_e$ vs. $\delta=\omega_{21}-\omega$ calculated from Eq.~(\ref{eq:bal}) for the same $n_s,T,\Omega_R$, $\gamma$, and $\omega$. The high- and low-$T_e$ branches are plotted by the solid lines, the dashed arrows indicate inter-branch switching.}
\end{figure}

The solution of Eq.~(\ref{eq:bal}) gives the stationary temperature of the electron system. A way of obtaining this solution graphically is illustrated in Fig.~{\ref{fig:4}}. The solution is defined by the intersection of a solid line and a dashed line, which represent the cooling rate and the heating rate, respectively. For the numerical evaluation, we assumed that $\nu_E$ is determined by collisions with He vapor atoms. However, we note that for $^3$He at $T=0.4$~K the energy relaxation rate can be affected by the interaction of electrons with short-wavelength ripplons, cf.~\cite{Platzman1999}. 

Important for the onset of bistability is that the spectral peak of resonant absorption rate $r(\omega)$ is narrow and that its position depends on $T_e$ in terms of $\Delta\omega_{21}(T_e)$. This leads to multiple solutions of Eq.~(\ref{eq:bal}) in a certain frequency range for sufficiently large MW power, as seen from Fig.~\ref{fig:4}~\cite{Ryvkine}. In particular, there exist two stable states (indicated by circles in Fig.~\ref{fig:4}): the one with comparatively high $T_e$ and another one with comparatively low $T_e$. The high-$T_e$ state ceases to exist for $\delta<\delta_1$, while the low-$T_e$ state is not possible for $\delta>\delta_2$. The boundaries $\delta_1$ and $\delta_2$ of the bistability interval depend on radiation power and shift towards negative values with increasing $\Omega_R$.

This theoretical model explains well the experimentally observed behavior. The calculated $T_e$ as a function of $\delta$ is shown in the inset of Fig.~\ref{fig:4}. Two stable solutions are plotted by the solid lines. As we increase $\delta$ starting from a large negative value, i.e. approach the resonance by sweeping $E_{\perp}$ from a low-field side, the system follows the low-$T_e$ branch as long as this solution exists. At the end of the branch, the system undergoes a transition to a new equilibrium state with higher $T_e$, as indicated by the upward arrow. Correspondingly, in the experiment we expect to observe an abrupt jump of the resonance response. On the other hand, when we decrease $\delta$ starting from a large positive value, i.e. sweep through the resonance from a high-$E_{\perp}$ side, the system remains on the high-$T_e$ branch until it ends. Then the system switches to the small-$T_e$ branch, as shown by the downward arrow. In the experiment, it could be seen as a sudden decrease of the resonant response. 

Our model assumes that the system is spatially uniform and the absorption linewidth $\gamma$ is independent of $T_e$. Heating-induced broadening of the absorption line might account for the difference in the response for $T=0.4$~K and 0.2~K (see Figs.~\ref{fig:1} and \ref{fig:3}). For $T=0.4$~K and for high $T_e$ the linewidth $\gamma$ becomes comparable to the inhomogeneous broadening, extending the frequency range where the system stays on the high-$T_e$ branch. At $T=0.2$~K this does not happen, $\gamma$ is much smaller, and the shape of the observed response in closer to the calculated one. A more detailed discussion will be provided elsewhere.

In summary, we have directly observed the long-sought intrinsic optical bistability in resonant inter-subband absorption of a quasi-2D electron liquid. The effect is due to the electron-electron interaction, which leads to strong spatial correlations in the electron liquid and as a consequence, to a strong shift of the inter-subband absorption line with varying distribution over the electron states. The bistability was studied by measuring the electron magnetoconductivity. The proposed theoretical model is in good agreement with the experimental results.

We acknowledge valuable discussions with P. M. Platzman and D. Ryvkine. DK and KK were supported in part by the MEXT, Grant-in-Aids for Scientific Research, MID was supported in part by the NSF, grant EMT/QIS 0829854, MJL was supported in part by the EPSRC.

%\bibliography{refdata}
%

\begin{widetext}
\newpage

\noindent \large\bf{SUPPLEMENTARY MATERIAL: the talk by M. I. Dykman at the International Meeting on ``Floating Electrons on Helium for Quantum Computing", Paris (2006)}

\includegraphics[scale=0.62]{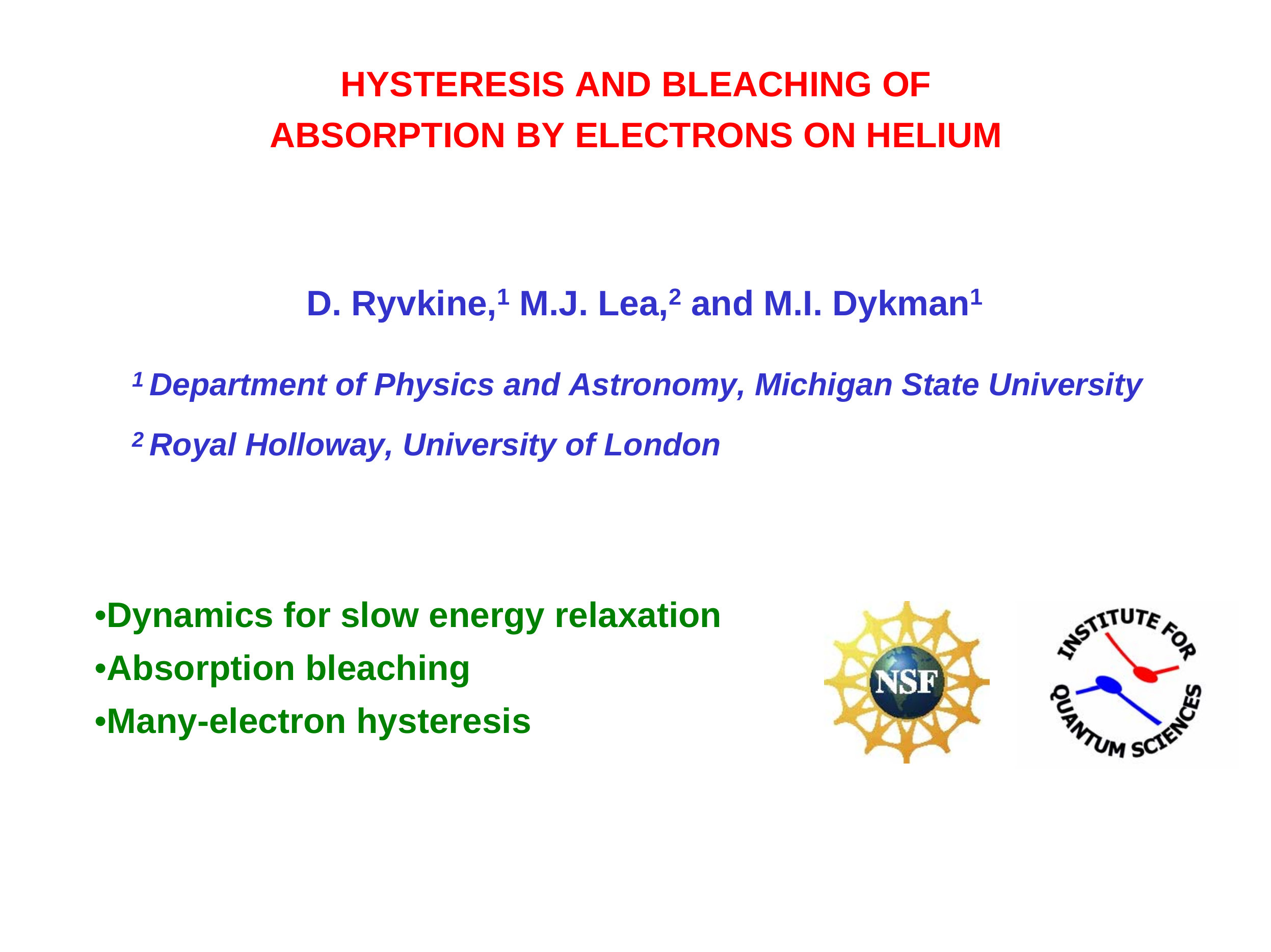}

http://www.princeton.edu/$\sim$lyon/Paris\_2006

\centering
\includegraphics[scale=0.62]{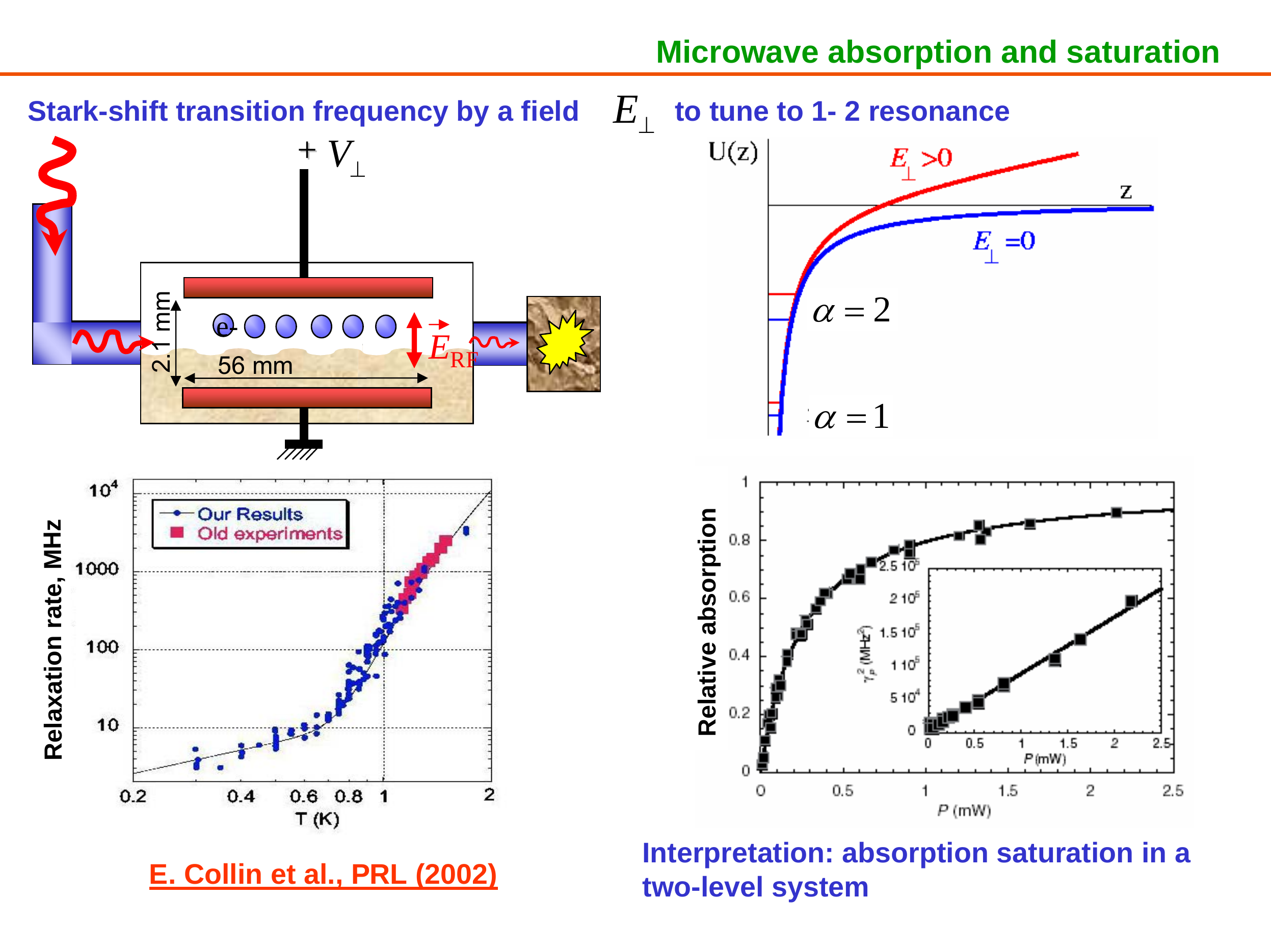}

\includegraphics[scale=0.62]{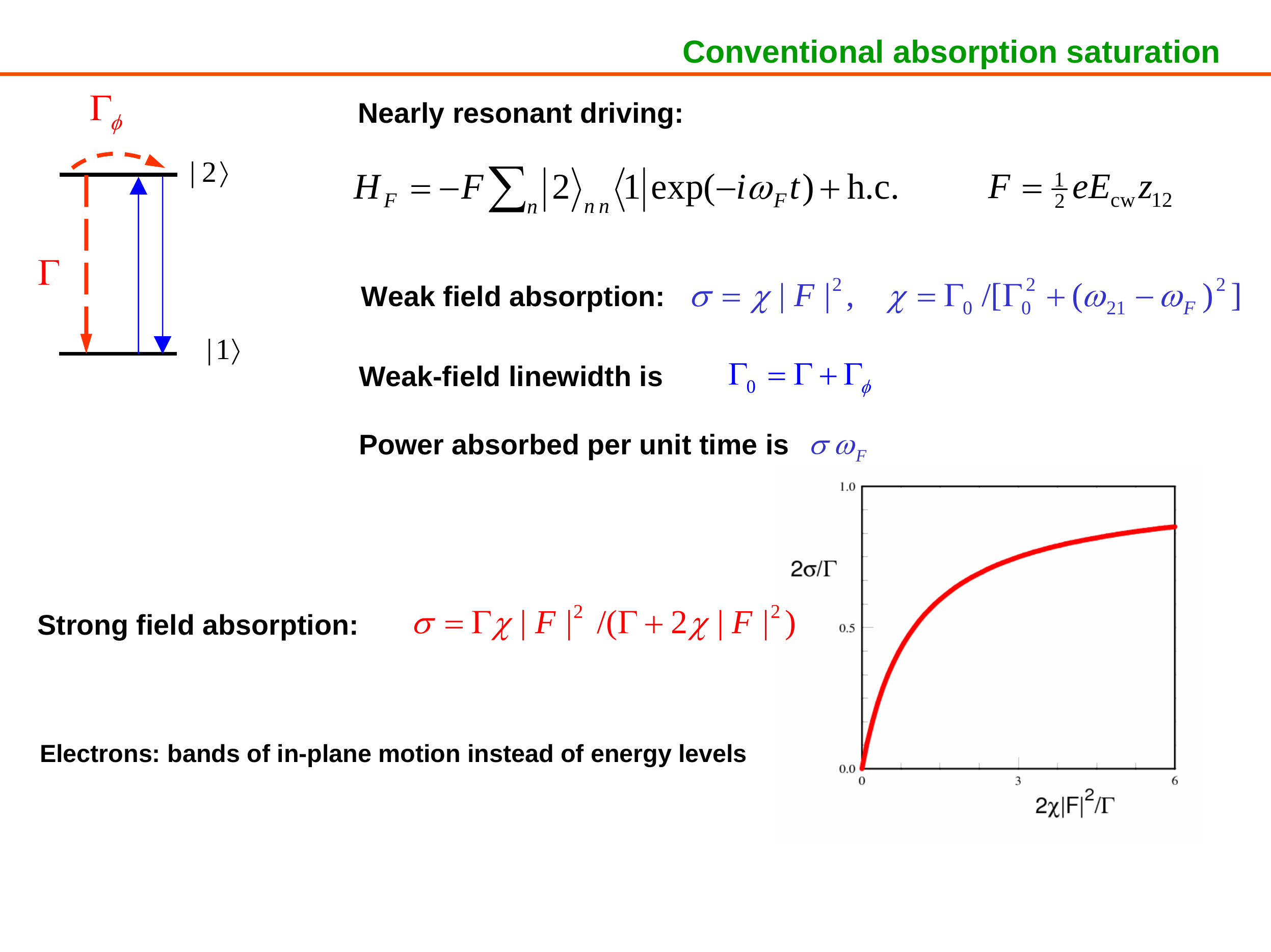}

\includegraphics[scale=0.62]{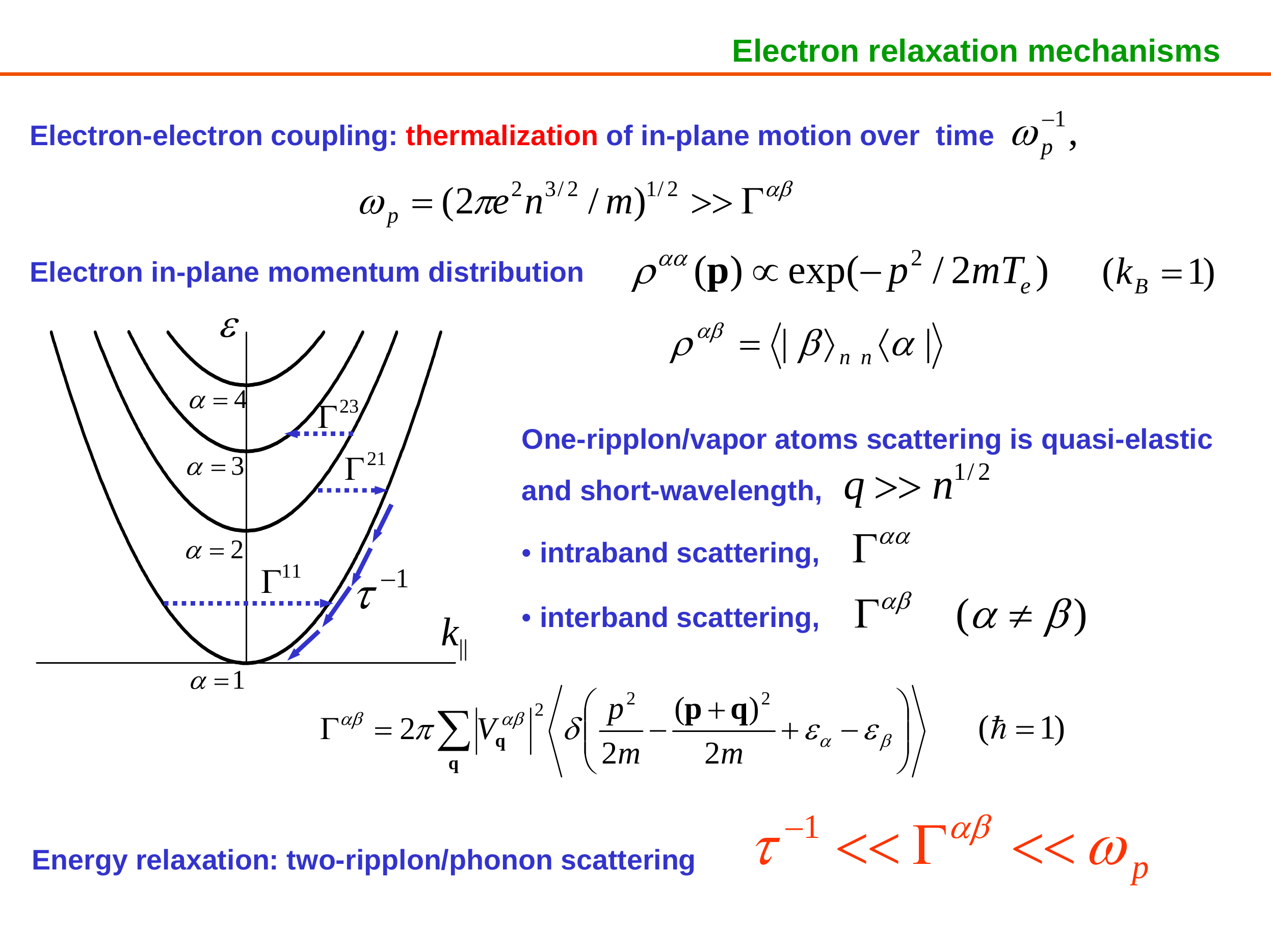}

\includegraphics[scale=0.62]{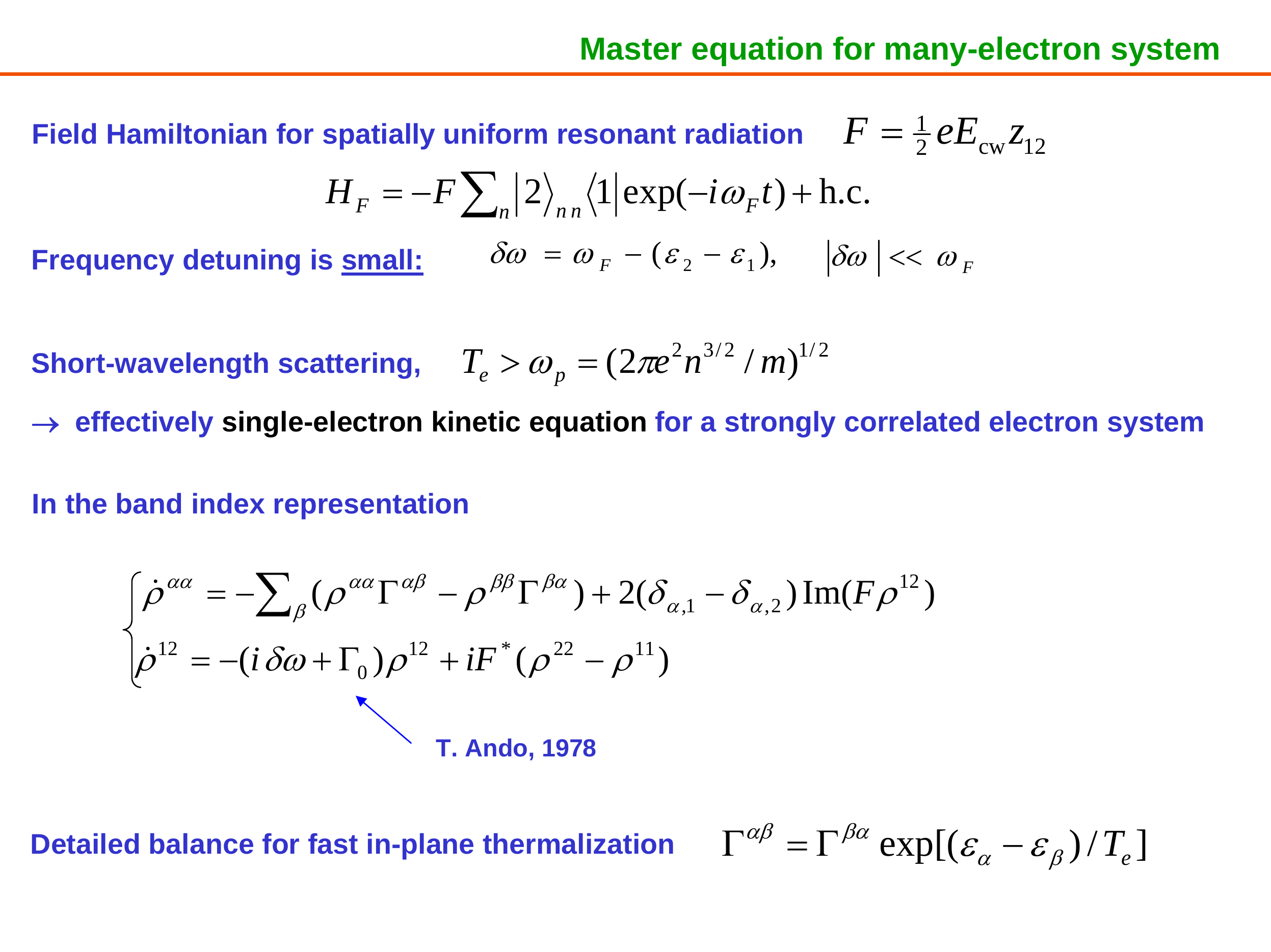}

\includegraphics[scale=0.62]{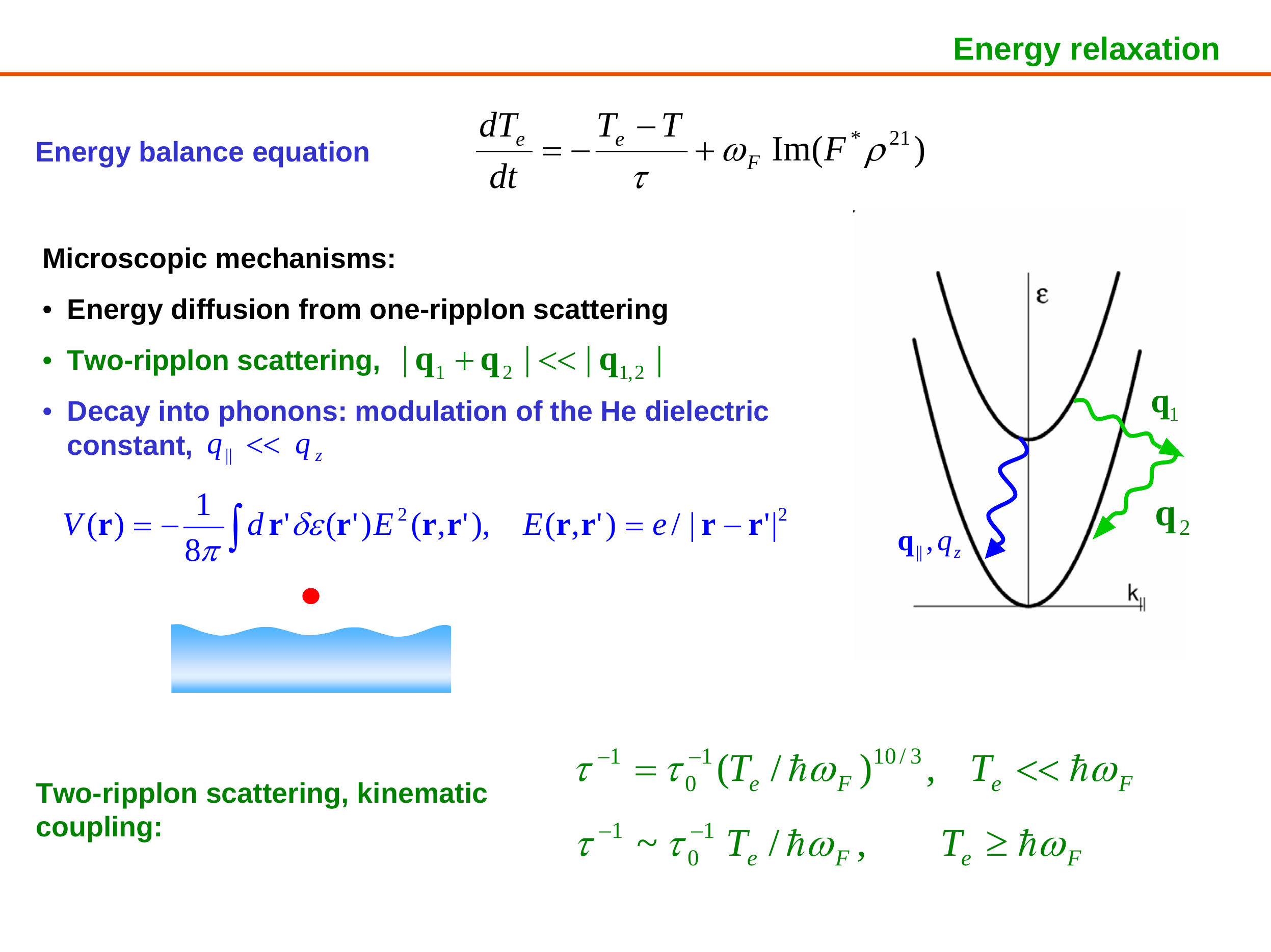}

\includegraphics[scale=0.62]{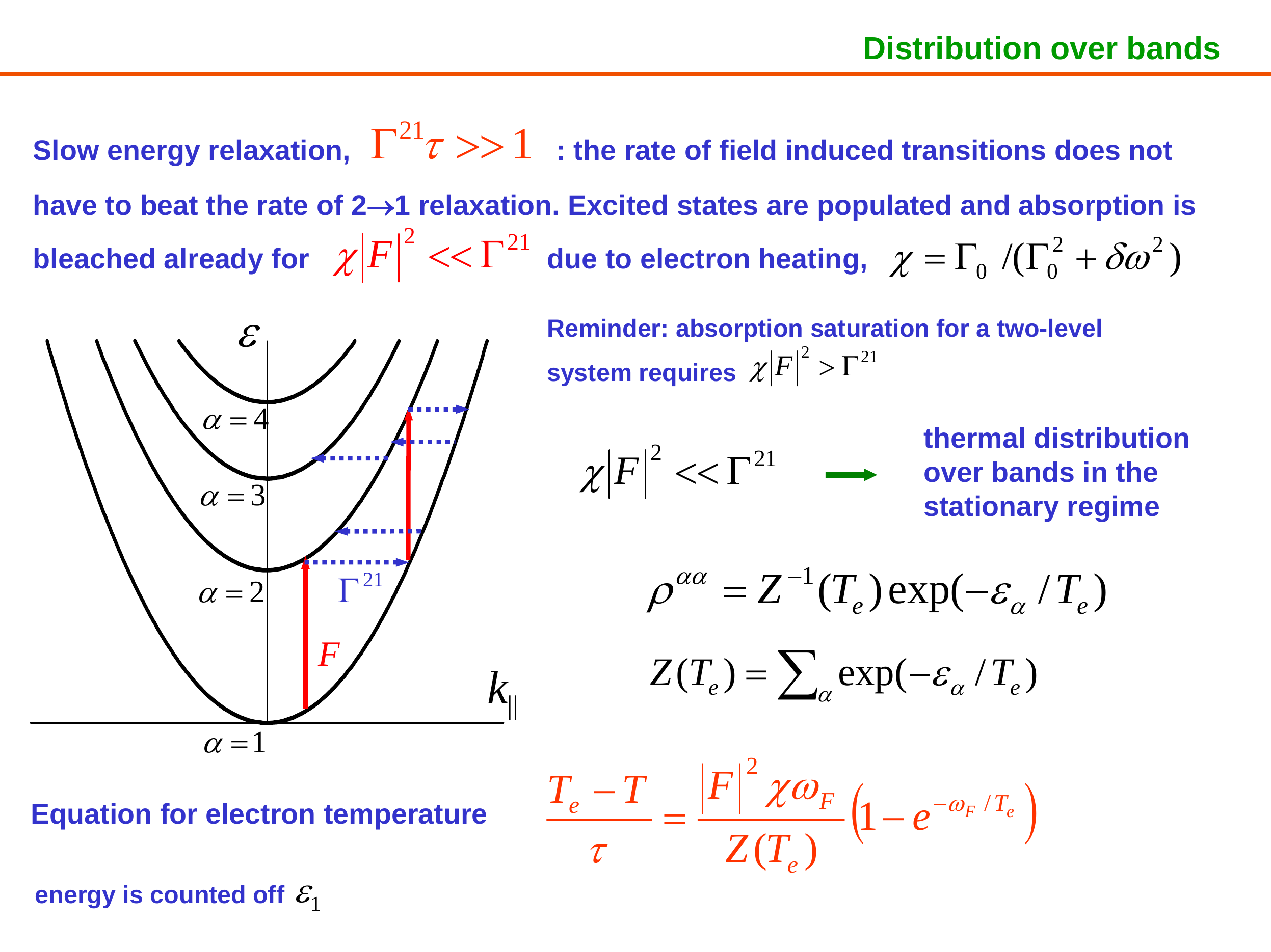}

\includegraphics[scale=0.62]{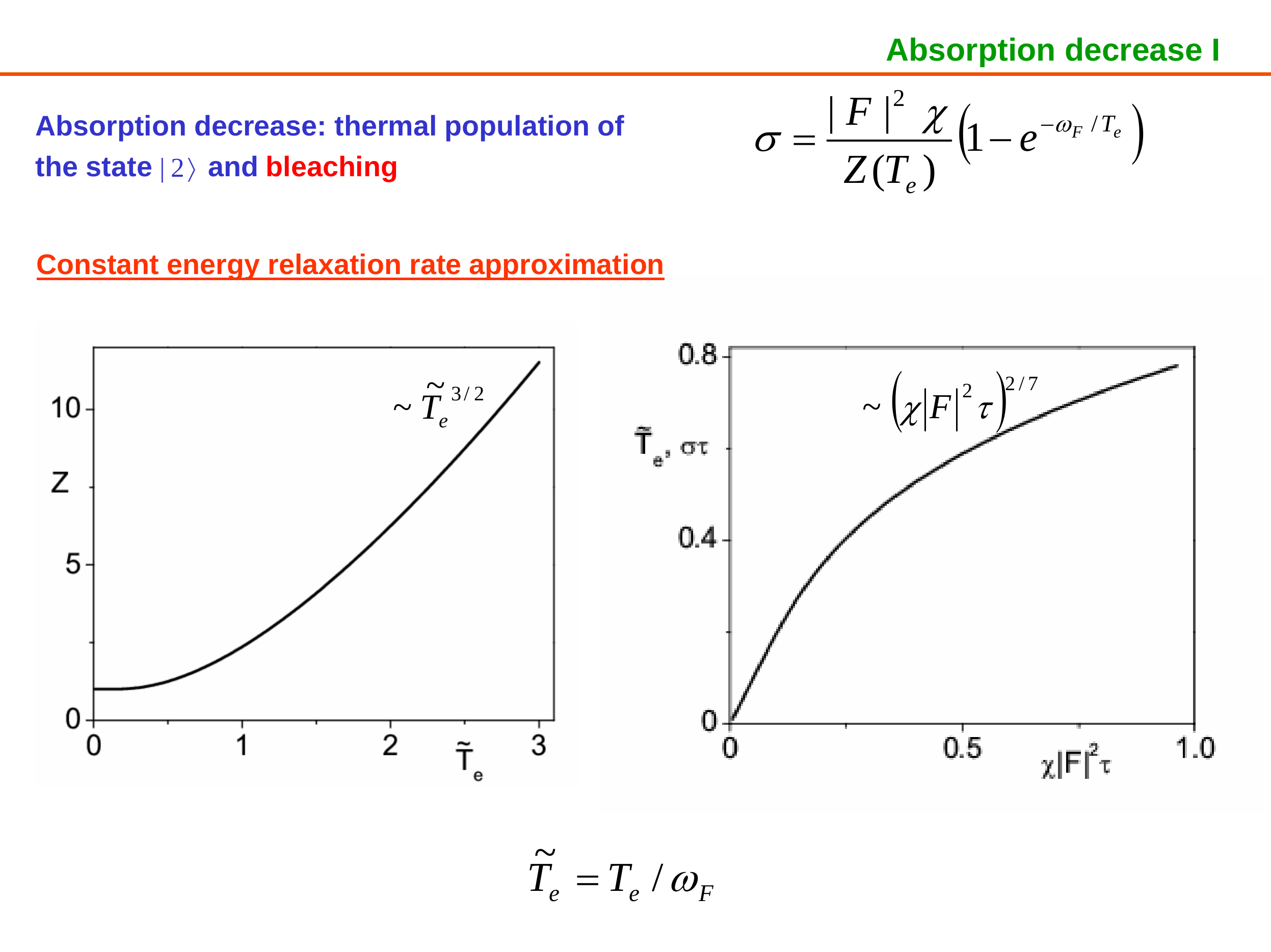}

\includegraphics[scale=0.62]{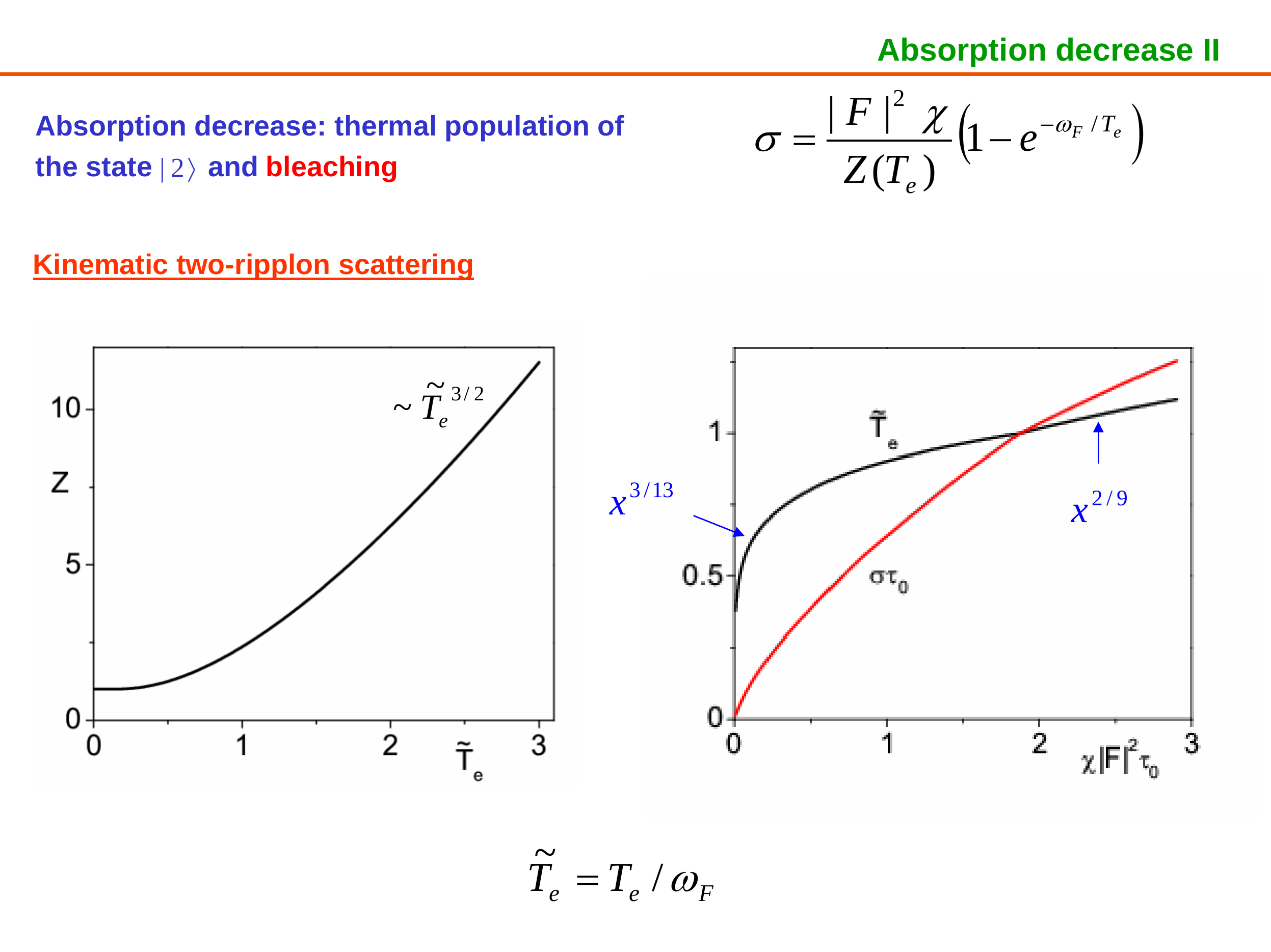}

\includegraphics[scale=0.62]{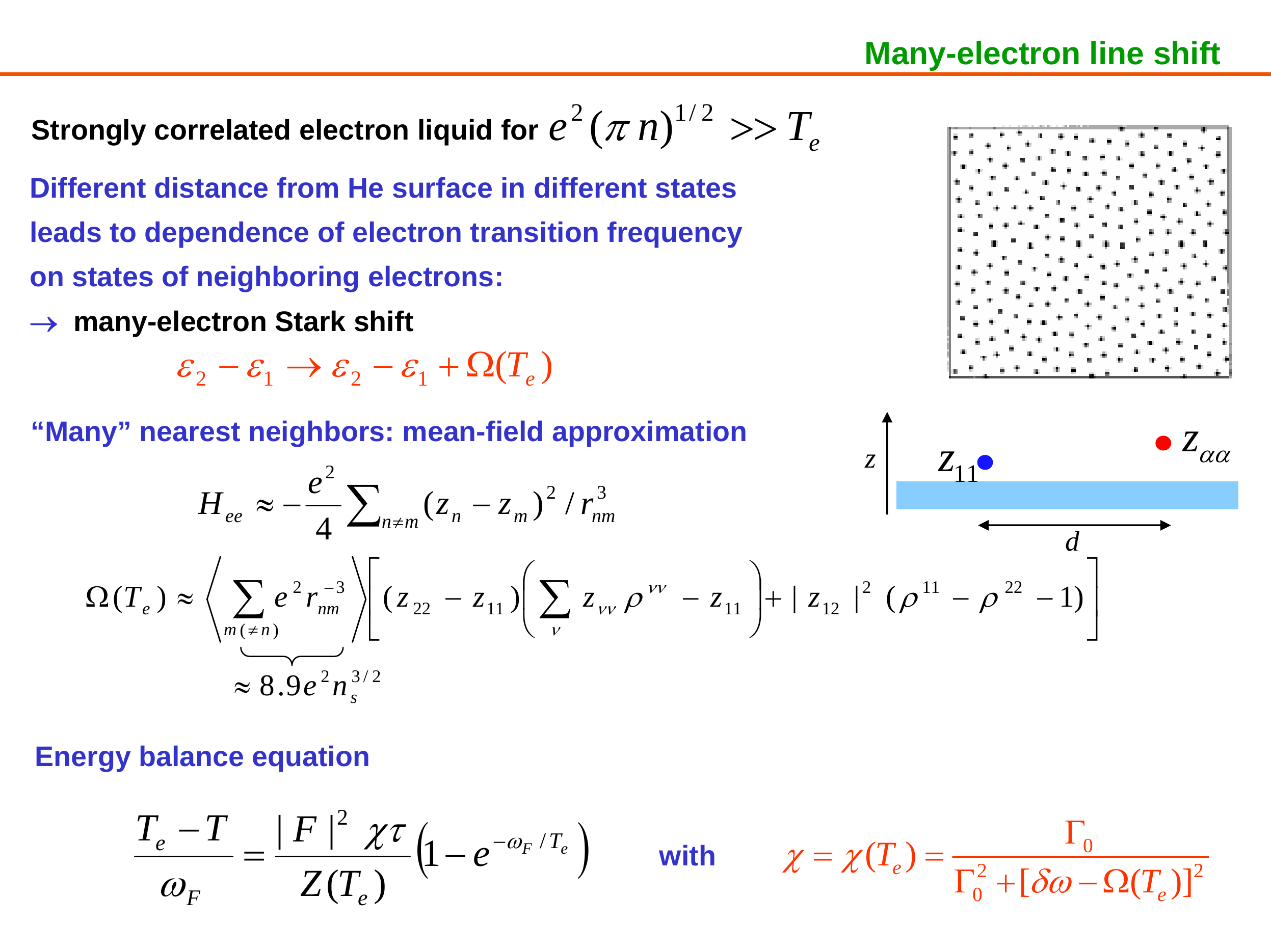}

\includegraphics[scale=0.62]{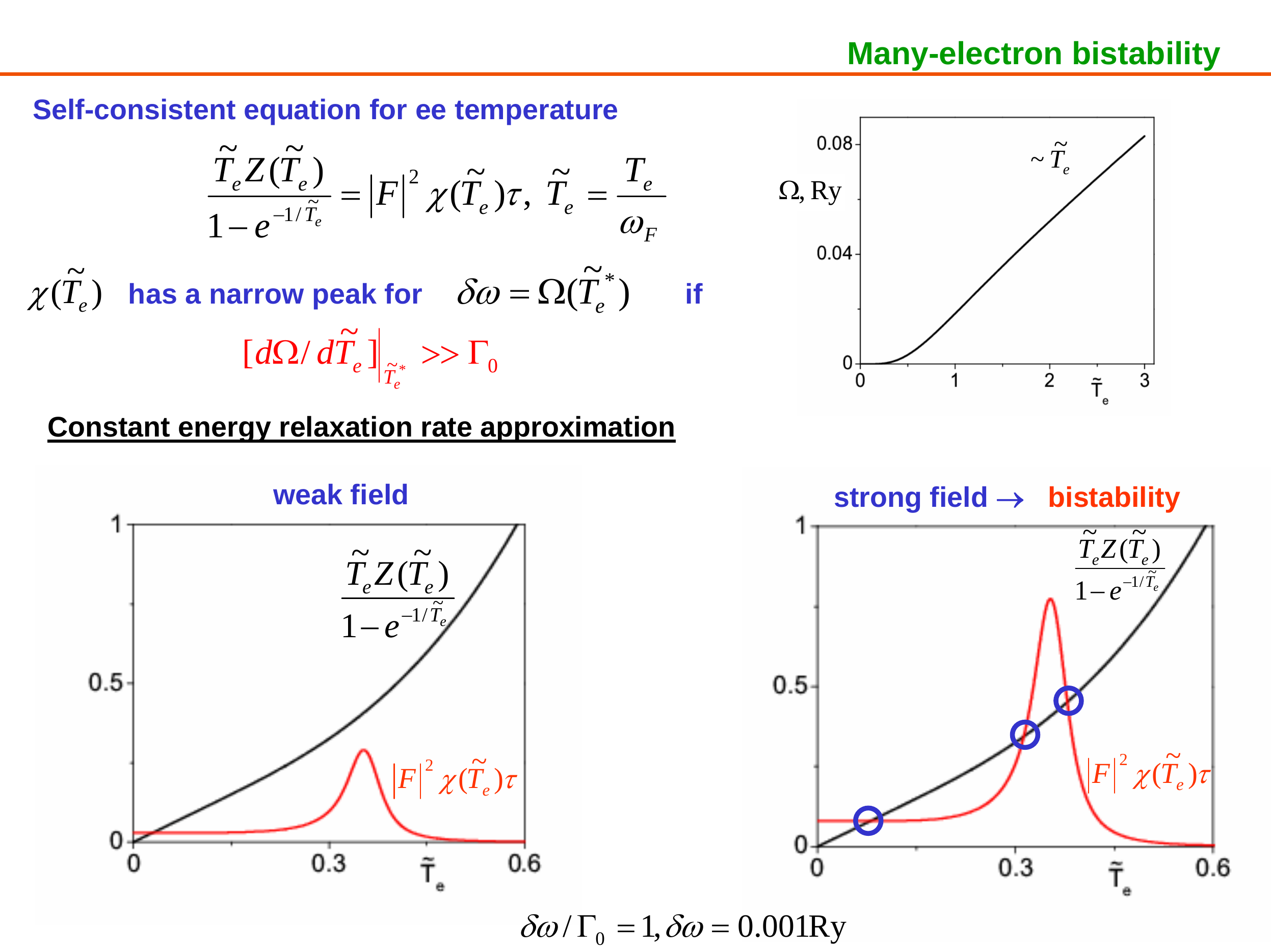}

\includegraphics[scale=0.62]{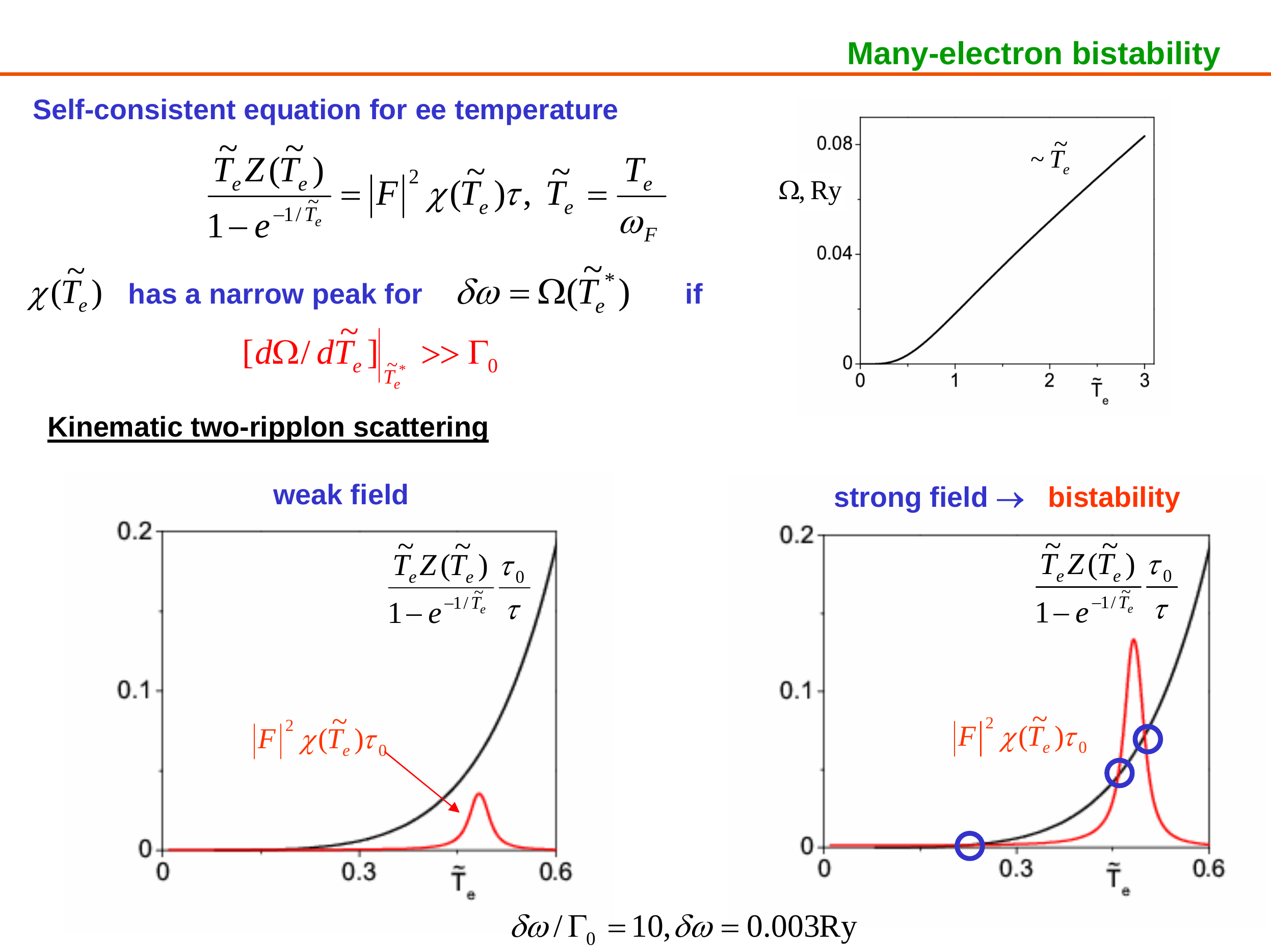}

\includegraphics[scale=0.62]{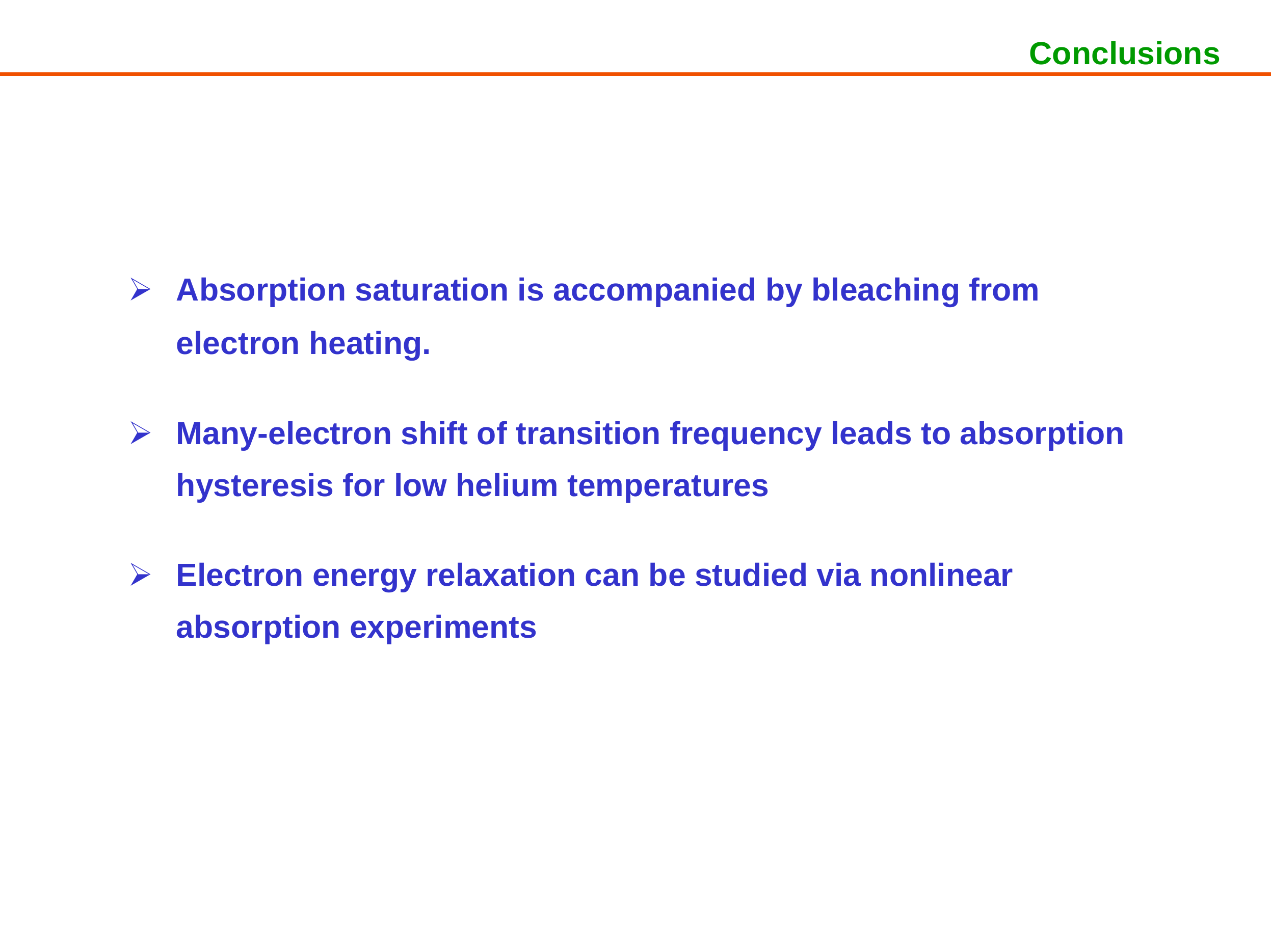}

\end{widetext}
\end{document}